\pdfoutput=1
\documentclass[reprint,aps,prd,superscriptaddress,showpacs,preprintnumbers,amsmath,amssymb,floatfix]{revtex4-1} 

\usepackage[euler]{textgreek}
\usepackage[mathlines]{lineno}
\usepackage{graphicx,subfigure}
\usepackage{epstopdf}
\usepackage{dcolumn}
\usepackage{bm}
\usepackage{color}
\usepackage{upgreek}
\usepackage{afterpage}
\usepackage{tikz, pgfplots}
\usepackage[version=4]{mhchem} 
\usepackage[colorlinks=true,unicode]{hyperref}

\newcommand{\cevns}{\ensuremath{\text{CE\textnu{}NS}}}

\pgfplotsset{compat=1.17} 

\begin{document}

\preprint{ \today}

\title{Neutron capture-induced silicon nuclear recoils for dark matter and \cevns 
}

\author{K.~Harris} \email{Corresponding author: kathryn.harris@ucdenver.edu}  \affiliation{Department of Physics, University of Colorado Denver, Denver, Colorado 80217, USA}
\author{A.~Gevorgian}  \affiliation{Department of Physics, University of Colorado Denver, Denver, Colorado 80217, USA}
\author{A.J.~Biffl}  \affiliation{Department of Physics, University of Colorado Denver, Denver, Colorado 80217, USA}
\author{A.N.~Villano} \email{Corresponding author: anthony.villano@ucdenver.edu}  \affiliation{Department of Physics, University of Colorado Denver, Denver, Colorado 80217, USA}

\smallskip
\date{\today}

\noaffiliation


\smallskip

\begin{abstract}
Following neutron capture in a material, there will be prompt nuclear recoils in addition to the
gamma cascade. The nuclear recoils that are left behind in materials are generally below 1\,keV
and therefore in the range of interest for dark matter experiments and \cevns\ studies --- both as
backgrounds and calibration opportunities. Here we obtain the spectrum of prompt nuclear recoils
following neutron capture for silicon.   

\end{abstract}


\maketitle

%
%
%
%
%

%
%
%
%
%
%
%

\section{\label{sec:intro}Introduction}
The residual nuclear recoils left after neutron capture have been used before to probe the details
of the slowing down of atoms in material~\cite{PhysRevA.11.1347,PhysRevD.103.122003}.  However,
the complications of the post-capture cascades and possible in-flight decays make the expected
energy of the residual nuclear recoils (NRs) nontrival to calculate. The energy of residual NRs
depends on the details of the capture cascade like the levels visited and the lifetimes of levels.
The work of Firestone~\cite{Firestone} in cataloging this information from experiments in prompt
neutron activation analysis (PGAA) is a key to being able to make the detailed NR energy
deposition models for silicon. 

Slowing-down models for the capture nuclei in their matrix are also a key component of correctly
doing the modeling. We use the approximation that nuclei that are slowing down do so with a
constant acceleration and we choose the acceleration to be in line with the Lindhard
model~\cite{osti_4701226}.  

Direct dark matter search experiments are often searching for low-energy NRs very near their
detector thresholds. The community has recently turned to neutron
capture~\cite{PhysRevD.106.032007,PhysRevD.103.122003,PhysRevD.105.083014} as a means to provide
very low-energy NRs near today's best detector thresholds --- below around 100\,eV in recoil
energy.  Similar efforts exist in the \cevns\ community~\cite{Thulliez_2021}. In both communities,
thermal neutron capture also exists as a potential background to the signal
events~\cite{biffl2022critical}. These studies show that a detailed understanding of the recoil
spectrum resulting from neutron capture is needed, and we provide that here for silicon detectors.  

\section{\label{sec:cascades}Postcapture Cascades}
For thermal neutron captures, each nuclear deexcitation releases approximately the neutron
separation energy, $S_n$, for the capturing isotope.  For intermediate and heavy nuclei, the
sequence of states that the residual nucleus passes through can be complex and have many emitted
gamma rays. This is the subject of long-standing data collection and modeling
efforts~\cite{Firestone}.

\begin{figure}[!htb]
   \includegraphics[width=\columnwidth]{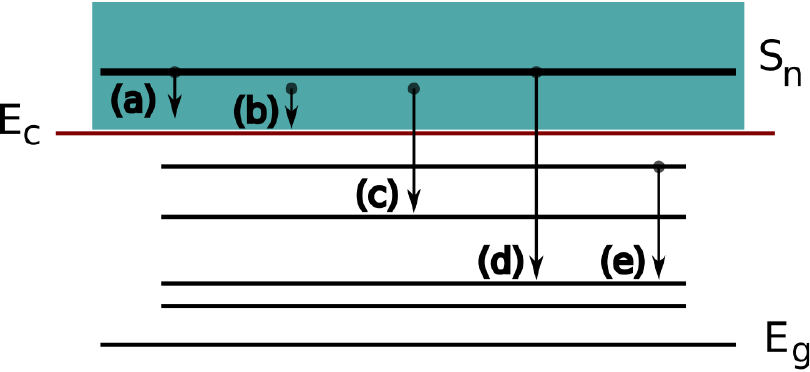}
   \caption{\label{fig:cascade}(Color online) The typical classification for capture cascade gamma
   rays.  Horizontal lines are various energy demarcations with $S_n$ representing the neutron
   separation energy, $E_g$ representing the ground-state energy, and $E_c$ representing the
   (arbitrary) cut off between discrete and continuum states.  Transitions are denoted by arrows
   and belong to one of five categories: (a) primary gammas with a final state in the
   continuum; (b) secondary gammas within the continuum; (c) secondary gammas from a continuum
   state to a discrete state; (d) primary gammas with a discrete final state; and (e) secondary
   gammas between two discrete states. 
   }
\end{figure}

The classification of individual deexcitations coming from the cascade has become standard and is
useful for relating the properties of the cascade to the nuclear structure.  In addition, this
classification aides in Monte Carlo codes to generate specific cascade realizations.  Generally, a
critical energy $E_c$ is chosen below the neutron separation energy such that nuclear levels below
are treated individually with their appropriate properties and levels between this energy and the
neutron separation energy are treated statistically. This breaks all released gamma rays into
several categories as displayed in Fig.~\ref{fig:cascade}: (a) primary to continuum; (b) continuum
to continuum; (c) continuum to discrete; (d) primary to discrete; and (e) discrete to discrete.  
 
In our treatment of silicon here we will take $E_c \simeq S_n$, that is, we will treat all
cascades as dicrete. This treatment may be difficult to implement for heavy nuclei as this
approximation is most accurate for nuclei with low masses.  Thus far, however, we have
successfully treated 27\% of germanium cascades in this fashion and are continuing to test this
treatment on the remaining 73\%.

\begin{figure}[!htb]
   \includegraphics[width=\columnwidth]{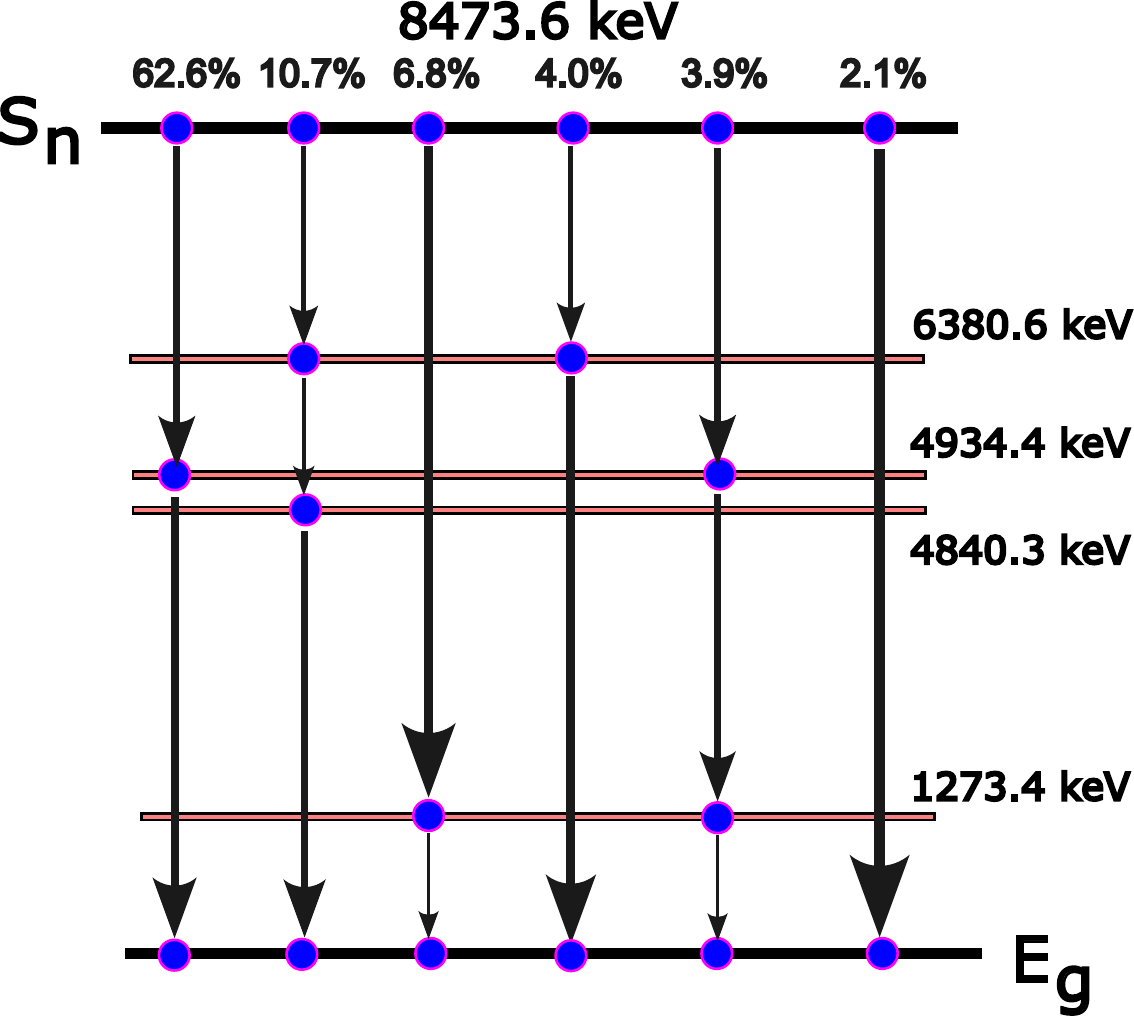}
   \caption{\label{fig:cascade-diagram}(Color online) A diagram of the six most prevalent cascades of
natural silicon. All of the cascades start with the capture of approximately thermal neutrons on
the nucleus $^{28}$Si at the neutron separation energy of the final nuclear state, $^{29}$Si.  
   }
\end{figure}

In PGAA measurements, it is easy to extract the prevalence of a specific gamma ray in the final
state per 100 captures. We prefer the slightly different organization of giving the probability of
a given \emph{cascade path}. The key publication we use to sort out the cascade probabilities is
the paper of Raman~\cite{PhysRevC.46.972}. Figure~\ref{fig:cascade-diagram} shows the cascade
paths of the six most probable cascades for a natural silicon composition. The cascades shown
there account for approximately 90\% of the total cascades. Some of the gamma rays appear in
multiple cascades so it is clear that the probability to find a specific gamma ray after
capture --- as is often measured --- is not quite the same information as the cascade probabilities that
we have compiled. 

\begin{figure}[!htb]
    \includegraphics[width=\columnwidth]{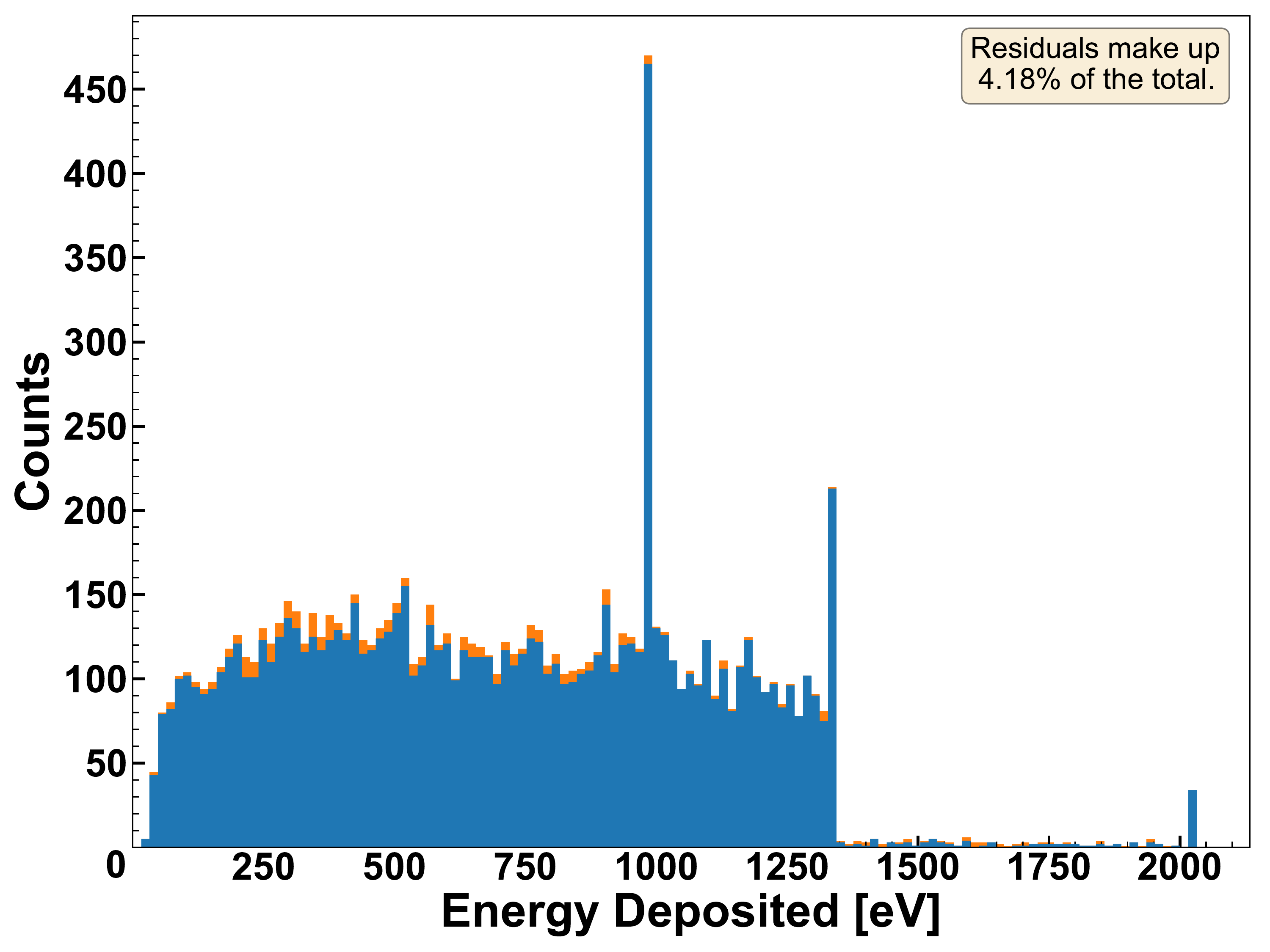}
    \caption{\label{fig:residuals}(Color online) A stacked histogram of energy deposits simulated
by \texttt{nrCascadeSim v1.5.0}~\cite{Villano2022}. On the bottom in blue (dark) are the cascades
listed in Table \ref{tab:acc_prob}. Stacked on top of them in orange (light) are the remaining
cascades.  }
\end{figure}

In our reorganization of this typical capture information, we have extracted the specific
cascades which account for 95.63\% of the total captures. This information is shown in
Table~\ref{tab:acc_prob} and is enough to construct an accurate model of the NRs left behind by
capturing thermal neutrons. Figure~\ref{fig:residuals} visualizes the effect the missing cascades
have on the full spectrum of energy deposition. We have also gathered in the table the half-lives
of each intermediate level where data are available and have otherwise used the Weisskopf
estimates~\cite{PhysRev.83.1073}.

Level lifetimes are important for our modeling because even in a dense (crystalline) matrix the
intermediate-state half-lives are typically short enough to allow for a ``decay in flight.'' In
other words, it is not a good approximation to assume that the excited nucleus in each
intermediate level stops before the subsequent decay. This has important implications for our
kinematics calculations later. We select a specific cascade path to model so we should technically
be using a level lifetime corrected for other branchings. The differences are small --- probably well
below 10\% --- because most of our highly-probable cascades involve the dominant decay branch.

\begin{table}[!hbt]
\begin{tabular}{c  c  c  c}
\hline
\hline 
	Isotope &  Probability (\%)   & Energy levels (keV) & Half-lives (fs) \\ \hline
	$^{28}$Si&	 62.6&	4934.39         &   0.84       \\
	$^{28}$Si&	 10.7&	6380.58, 4840.34&	0.36, 3.5  \\
	$^{28}$Si&	 6.8 &	1273.37         &	291.0      \\
	$^{28}$Si&	 4.0 &	6380.58         &	0.36       \\
	$^{28}$Si&	 3.9 &	4934.39, 1273.37&	0.84, 291.0\\
	$^{28}$Si&	 2.1 &	$\cdots$              &	$\cdots$         \\
	$^{29}$Si&	 1.5 &	6744.10        &	14         \\
	$^{30}$Si&	 1.4 &	3532.9, 752.20 &	6.9, 530   \\
	$^{29}$Si&	 1.2 &	7507.8, 2235.30&	24, 215    \\
	$^{29}$Si&	 0.4 &	8163.20        &	w(E1)[0.0019]   \\
	$^{30}$Si&	 0.4 &	5281.4, 752.20 &	w(E1)[0.0069], 530 \\
	$^{29}$Si&	 0.3 &	$\cdots$               &	$\cdots$          \\
	$^{30}$Si&	 0.3 &	4382.4, 752.20 &	w(E1)[0.012], 530 \\
	$^{30}$Si&	 0.03 &	$\cdots$               &	$\cdots$          \\
\hline
\hline
\end{tabular}
   \caption{\label{tab:acc_prob}
	A table displaying the probability of each cascade.  This table includes only the cascades
used for our model.  The isotope listed is the isotope on which the neutron captures; the energy
levels and half-lives are therefore for an isotope of silicon with one more neutron.  A half-life
entry in [brackets] preceeded by w(E1) specifies that the half-life is unknown and the Weisskopf estimate for an electric
dipole transition was used~\cite{PhysRev.83.1073}.  
}
\end{table}

The possibility of ``decay in flight'' also makes a calculation of the slowing-down of recoiling
atoms germane to our modeling. Here, we use a constant acceleration to model this slowing down,
consistent with the average stopping power derived by Lindhard~\cite{osti_4701226}. Lindhard used
a generic Thomas-Fermi potential for all ions, and the result was a stopping power (acceleration) $S$
that depended on energy for slow nuclei between about 100\,eV and 1\,keV. We use the average of
that curve between those energies, $S=0.1$. 

To estimate a rough impact of the use of an average stopping power, we compared data generated by
\texttt{nrCascadeSim v1.5.0}~\cite{Villano2022} with stopping powers of $S=0.05$ and $S=0.15$ and
found the average difference to be $22\%$, with the 984\,eV peak differing by $97\%$. The
qualitative differences between the $S=0.05$ and $S=0.15$ stopping powers are minimal; the
greater stopping power results in larger peaks associated with full stops before decay, but covers
the same region with similar distributions for nonpeak events. While peaks are taller for the
greater stopping power, they are still noticeable in both cases. This indicates that one very
sensitive measure of the average stopping power in our energy spectrum is the ratio of the tallest
peak to the flat region.

\section{\label{sec:twostep}Two-Step Cascades}

For cascades which emit either one or two gamma rays (one- or two-step cascades), we were able to
analytically construct the distribution of total NR energies. This distribution will be what is
observed in a detector that experiences a neutron capture when all the gamma rays leave without
energy deposit. 

For one-step cascades, a single gamma is emitted back-to-back with the NR. The gamma energy in
this case is approximately the neutron separation energy, $E_{\gamma} \simeq S_n$. The NR energy,
$T$, is given approximately by
\begin{equation}
T \simeq \frac{S_n^2}{2M_A}.
\end{equation}
where $M_A$ is the mass of the recoiling nucleus. 

The two-step cascades are considerably more complex because of the possibility of decay in flight.
We work with a separation of the nuclear energy deposits into the first and second steps like:
$D_t = D_1 + D_2$. $D_t$ is the total deposited energy by the NRs. $D_1$ and $D_2$ are the
energies deposited before the intermediate decay and after the intermediate decay respectively.
The two other key variables we will use are the decay time, $t$, and the center-of-mass angle for
the decay, $\beta_{\mathrm{cm}}$. The energy deposits are deterministic functions of the decay
times and angles, both of which are in turn probabilistic random variables. 

The decay time represents how long it takes for the (instantaneously generated) intermediate state
to decay to the ground state and is exponentially distributed with the probability density
function (PDF) in Eq.~(\ref{eq:tdist}). The cosine of $\beta_{\mathrm{cm}}$ is assumed to be
uniformly distributed over $(-1,1)$ in the center-of-mass frame for the decay. The possible
correlation between gamma directions is mostly erased by the interaction of the excited state with
the lattice. We estimate that around 8$\times$10$^{-4}$\% of the time there could be a cascade
that emits two gammas nearly simultaneously --- in that case any correlation will remain but has not
been accounted for here, 

\begin{equation}\label{eq:tdist}
f(t) = \frac{\ln{2}}{t_{1/2}} \exp\left [-t\frac{\ln{2}}{t_{1/2}}\right ],
\end{equation}

The quantity $D_1$ can be expressed as a simple function of $t$, given that the recoiling nucleus
slows down with a constant (negative) acceleration, $a$,

\begin{equation}\label{eq:d1}
D_1(t) = T_1 - \frac{M_A^{\ast}(v_0-at)^2}{2},
\end{equation}
where $T_1$ is the total kinetic energy the intermediate state receives from the first gamma
recoil, $M_A^{\ast}$ is the mass of the intermediate state, and $v_0 = (2T_1/M_A^{\ast})^{1/2}$ is
the initial velocity.

Modeling the process with a fixed acceleration gives a unique stopping time, $t_s$. The
distribution of $D_1$ has a singular value of $T_1$ if $t > t_s$, but the PDF for $t < t_s$
depends on Eq.~(\ref{eq:tdist}) with a change of variables to the $D_1$ of Eq.~(\ref{eq:d1}). The
result is

\begin{widetext}
\begin{equation}
    \centering
    \begin{aligned}
        \label{eq:gd1}
        g(D_1)&=
            \begin{cases}
                g^0(D_1)+\exp{\left[ - \ln{2}\frac{t_s}{t_{1/2}}\right]} \delta(D_1-T_1) & ; D_1 \leq T_1\\
                0 & ; D_1 > T_1 
            \end{cases} \\
        g^0(D_1) &= \frac{\ln{2}}{|a|t_{1/2}\sqrt{2M_A^{\ast}(T_1-D_1)}}\exp{\left [-\ln{2}
        \frac{v_0-\sqrt{2(T_1-D_1)/M_A^{\ast}}}{|a|t_{1/2}} \right ]},
    \end{aligned}
\end{equation}
\end{widetext}
where $\delta(\cdot)$ is the Dirac delta function.

The PDF of $D_2$ is clearly dependent on $D_1$ because the value of $D_2$ depends on $t$ which is
deterministically related to $D_1$. We deal with this by using the joint PDF of $D_1$ and $D_2$,
$g(D_1,D_2)$. The PDF for $D_2$ can then be obtained by integrating the joint PDF over all $D_1$.

To obtain the joint distribution $g(D_1,D_2)$ we use a basic relationship from conditional
probability,

\begin{equation}
\label{eq:condprob}
g(D_1,D_2) = h(D_2|D_1)g(D_1).
\end{equation}
The PDF $h(D_2|D_1)$ is the PDF of $D_2$ given a specific value of $D_1$. The function $h$ is not
challenging to obtain because the only relevant random variable is $\beta_{\mathrm{cm}}$-- $t$ is
fixed because $D_1$ is fixed. We can then think of $D_2$ as a deterministic function of $D_1$ and
$\beta{\mathrm{cm}}$ like $D_2(t(D_1),\beta_{\mathrm{cm}})$.

\begin{figure}[!htb]
   \includegraphics[width=\columnwidth]{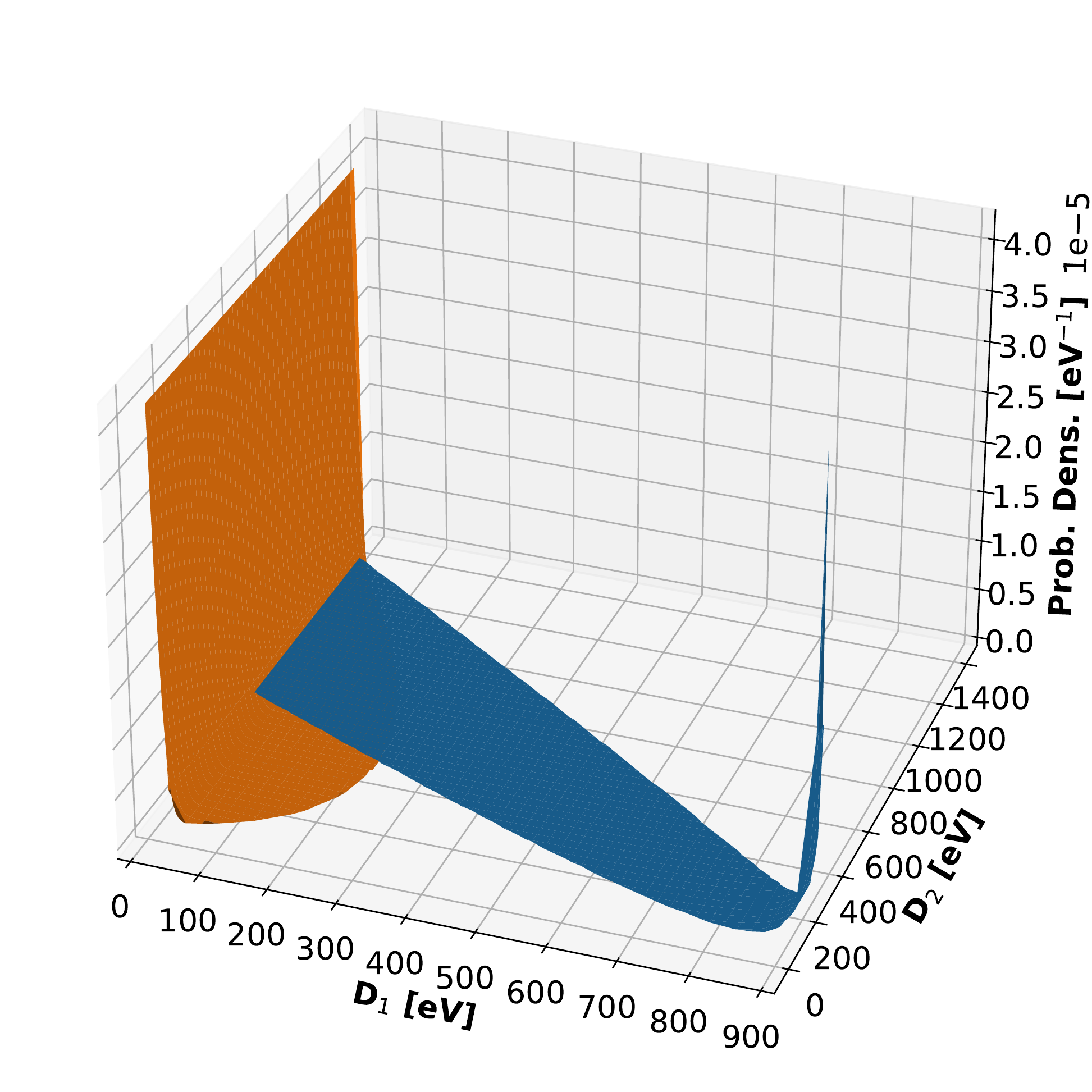}
   \caption{\label{fig:twostepjoint}(Color online) The two-dimensional joint PDF from
Eq.~(\ref{eq:condprob}). $D_1$ and $D_2$ are the energies deposited from the first and second
cascade step respectively. The darker (blue) surface is for a two-step cascade stopping at the
first excited state of $^{29}$Si; the lighter (orange) surface is for a two-step cascade stopping
at the (tenth) excited state of $^{29}$Si --- the most likely cascade.  
}
\end{figure}

To obtain the function $h$ we note that whatever kinetic energy the final nucleus has after the
intermediate decay will be the deposited energy~\footnote{we are not accounting for the
possibility that the recoiling nuclei escape the medium because their ranges are very small}. We
calculated bounds on this kinetic energy, $T_2$, given the value of $\beta_{\mathrm{cm}}$,

\begin{equation}\label{eq:d2minmax}
\begin{split}
T_2 = \frac{\Delta^2}{2M_A} &\left [ \frac{2M_A(T_1-D_1)}{\Delta^2} \right. \\
& \left. + 2 \sqrt{\frac{2M_A(T_1-D_1)}{\Delta^2}}\cos{\beta_{\mathrm{cm}}} + 1   \right ],
\end{split}
\end{equation}
where $\Delta$ is the difference between the intermediate state energy and the ground state.  The
value of $\cos{\beta_{\mathrm{cm}}}$ is between -1 and 1, so this gives a clear minimum and
maximum for this kinetic energy. An alternate form for the kinetic energy $T_2$ is given in
Appendix~\ref{app:rearrange}.  The total energy deposited can be zero if and only if the decay is
immediate and $\Delta$ is exactly halfway between the ground state and the neutron separation
energy $S_n$. The function $h$ is then
 
\begin{equation}
\label{eq:h}
h(D_2|D_1)= \\
    \begin{cases}
         \frac{1}{2\Delta}\sqrt{\frac{M_A}{2}\frac{1}{T_1-D_1}}& ; T_{2,\mathrm{min}} \leq D_2 \leq
T_{2,\mathrm{max}}\\
        0 & ; \mathrm{otherwise} 
    \end{cases} \\
\end{equation}
Using Eq.~(\ref{eq:h}) we constructed the joint distribution from Eq.~(\ref{eq:condprob}). The joint
distribution is plotted in Fig.~\ref{fig:twostepjoint} for two cascades --- one (orange) with a very
fast intermediate decay and another (blue) with a slower intermediate decay. The spike shown in
the slow distribution is a two-dimensional Dirac delta function that corresponds to the situation
when the intermediate recoil stops before the subsequent decay. In that case the values of $D_1$
and $D_2$ are fixed and are the values you would expect from at-rest one-gamma decay of each
excited nuclear state. The fast intermediate decay produces a behavior where $D_1$ tends to be
lower and increases toward zero.
With the joint distribution specified, the distribution of the total energy deposit, $D_t$ is 
obtained by the following integral:

\begin{equation}\label{eq:dtdist}
p(D_t) = \int_{D_1} g(D_1,D_t - D_1) dD_1.
\end{equation}

\begin{figure}[!htb]
   \includegraphics[width=\columnwidth]{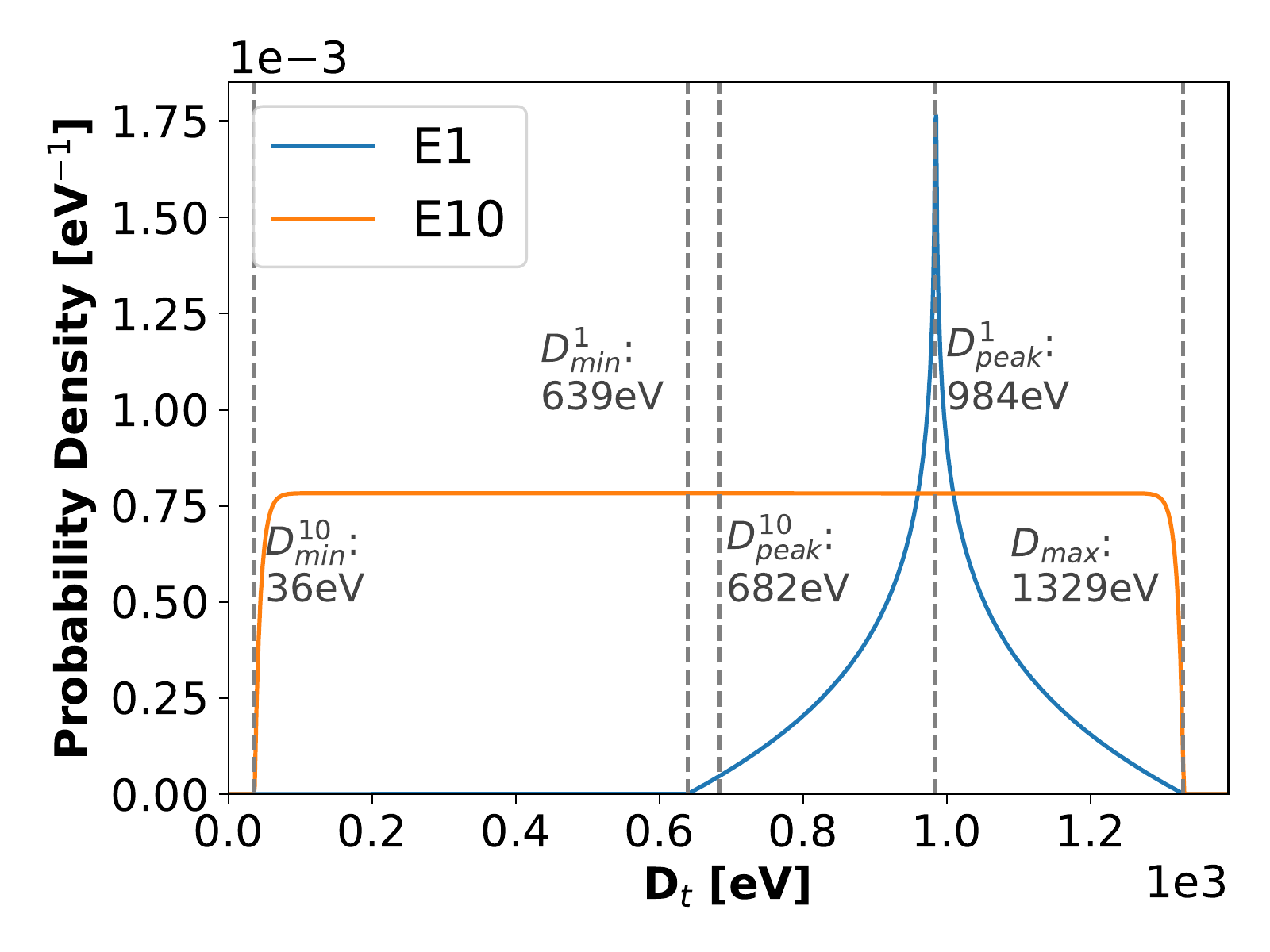}
   \caption{\label{fig:dtotpdf}(Color online) The PDF of the total deposited energy from nuclear
recoils, $D_t$, for this two-step cascade. E1 indicates the intermediate level being the first
excited state of $^{29}$Si; E10 indicates the intermediate level being tenth excited state of
$^{29}$Si --- the most likely cascade for Si.  
}
\end{figure}

The total distribution for $D_t$ is plotted in Fig.~\ref{fig:dtotpdf} for both example cascades.
Once again the ``spike'' in the slow cascade corresponds to the case where the intermediate recoil
stops before the subsequent decay --- that results in a fixed total energy deposited. This spike is
proportional to the Dirac delta function, and so cannot be shown on the correct scale. However, it
is easy to see the $D_t$ value for the spike and it must account for the remaining probability
after removing the integral of the plotted distribution. In the fast cascade most remnants of the
monoenergetic ``spike'' are gone and the total energy is nearly uniform between two fixed bounds.

\section{\label{sec:mc}Monte Carlo Approach}

For cascades with more than two steps we have not worked out the analytical distributions as we
have in Sec.~\ref{sec:twostep}. For these many step cascades we use a Monte Carlo approach that
allows us to include arbitrarily many steps in the sequence. The main limitation is that we
compute one thermal neutron capture realization at a time so that it might be prohibitive to
produce a PDF with sufficient smoothness (high statistics) for cascades with small overall
likelihood. On the other hand, those cascades represent only a small change to an experimental
spectrum~\cite{PhysRevD.105.083014}.

Using the information from Table~\ref{tab:acc_prob}, these steps are followed to generate one
Monte Carlo capture event:

\begin{enumerate}
\item Select a cascade with a probability based on the prevalence of that specific deexcitation
path.
\item Randomly select a decay time for the first intermediate state based on an exponential
distribution with the appropriate lifetime. This variable is $t$ from the two-step calculation. 
\item Calculate the first energy deposit $D_1$ based on the decay time and the stopping
acceleration. This is the slowing-down energy deposited in time $t$.
\item Adjust the kinetic energy of the recoil based on the kinematics of in-flight decay. This
adjustment is based on the center-of-mass angle of the emitted gamma, $\beta_{\mathrm{cm}}$ from
the two-step calculation.
\item Repeat steps 2--4 for each intermediate level. 
\end{enumerate}
The result of this procedure is a set of energy deposits $\{D_i\}$ with the same number of
elements as gamma rays emitted. The $D_i$s are saved and may be summed to obtain the total
deposited energy. The emitted gamma ray energies for each step are saved alongside the $D_i$s.
These steps are implemented in our public code~\cite{Villano2022}. 

\begin{figure}[!htb]
   \includegraphics[width=\columnwidth]{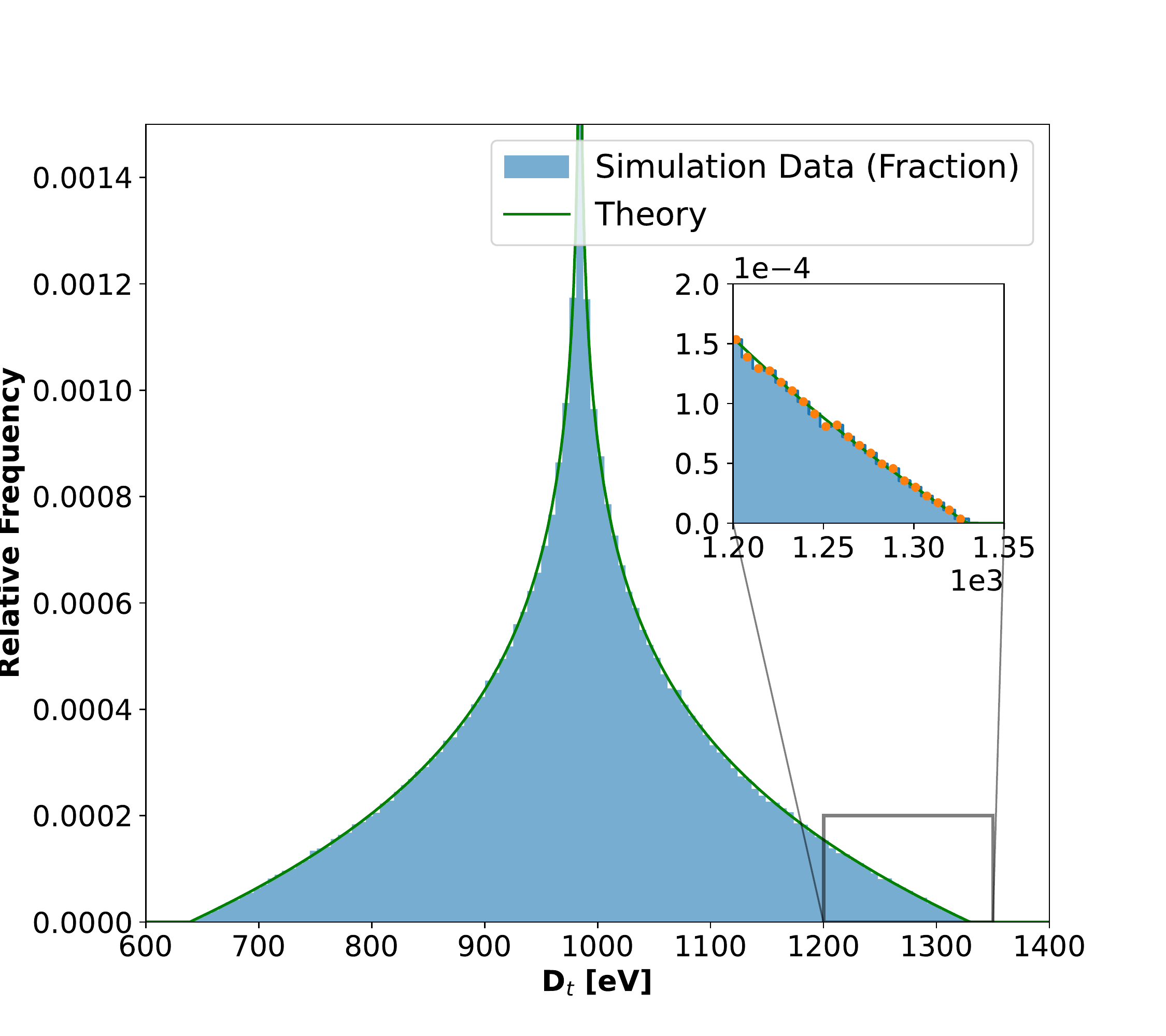}
   \caption{\label{fig:twostepcompare}(Color online) A comparison of the analytical PDF for the
two-step cascade (see Sec.~\ref{sec:twostep}) with that of the Monte Carlo procedure. The
histogram is the PDF derived from many events generated by the Monte Carlo procedure.  
}
\end{figure}

Figure~\ref{fig:twostepcompare} shows how the analytical calculation for the deposited energy of a
two-step cascade compares to our Monte Carlo procedure.  The two are an excellent match and the
reduced $\chi^2$ statistic is 1.02. 

\begin{figure}[!htb]
   \includegraphics[width=\columnwidth]{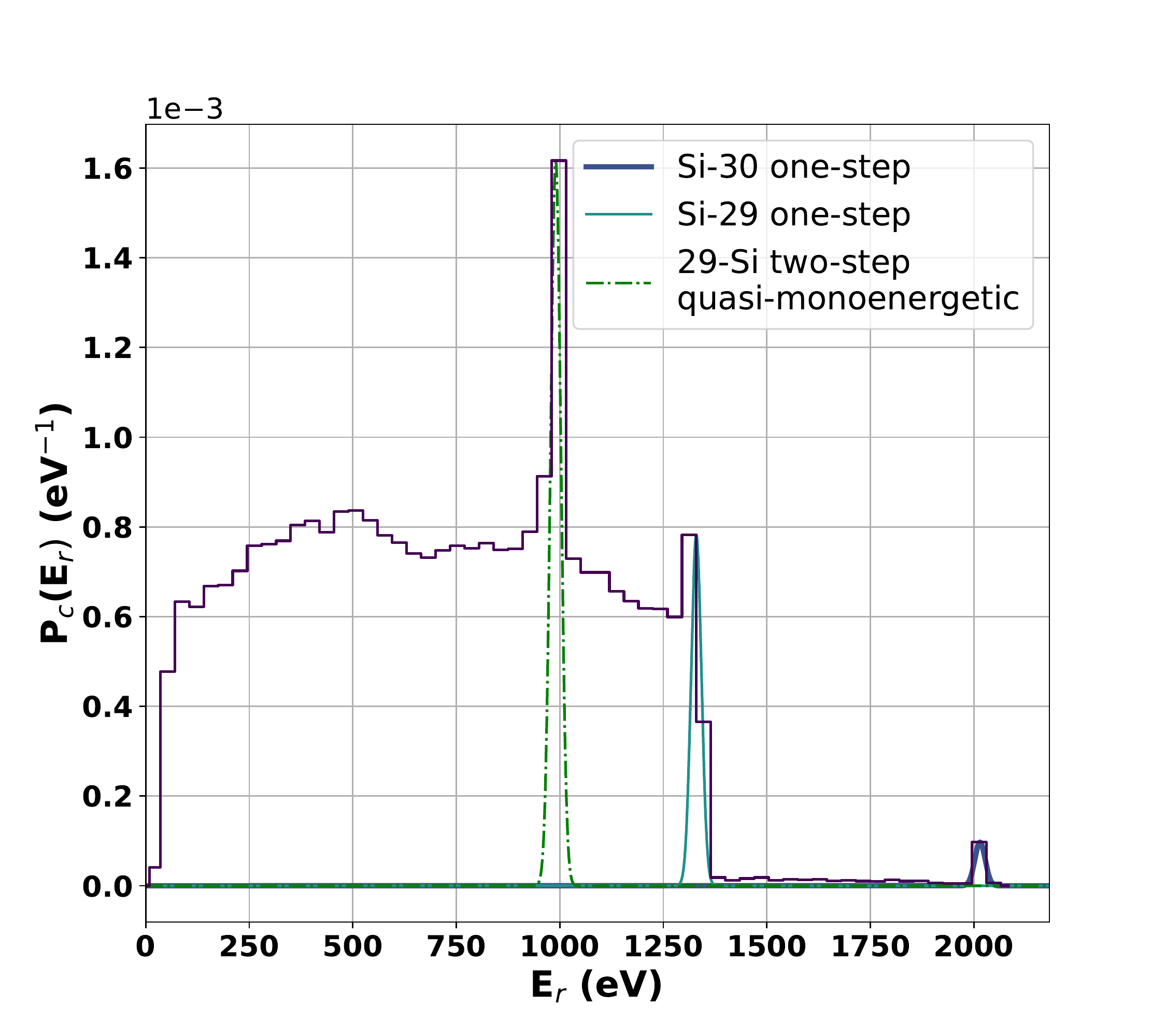}
   \caption{\label{fig:fullcapspec-resolution}(Color online) The complete silicon capture spectrum
using the data from Table~\ref{tab:acc_prob}. 95.63\% of all cascades are taken into account here.
A nominal 10\,eV resolution has been applied.  
}
\end{figure}

The full spectrum from all the cascades in Table~\ref{tab:acc_prob} is shown in
Fig.~\ref{fig:fullcapspec-resolution} with a 10\,eV nominal resolution applied. The sharpest peaks
come from the direct-to-ground transitions of $^{29}$Si and $^{30}$Si and the 6.8\% two-step
transition. In the two-step transition there is a sizeable probability of having the first NR stop
before the subsequent decay --- it is a quasimonoenergetic transition for this reason. 
When we use these spectra we account for the events that will be distorted by energy deposits from
the exiting gammas~\cite{PhysRevD.105.083014}. This is sometimes done by estimating what fraction
of cascades have gammas that escape (around 90\% for a cylindrical silicon detector of diameter
100\,mm and thickness 30\,mm). Other times we use a particle transport code like \texttt{Geant4}
to find exactly which cascade realizations have exiting gamma interactions. When using materials
that are more electron dense than silicon the fraction of interactions from exiting gammas
increases. It is typically true that those events that have interactions from exiting
(high-energy) gammas will be removed from the low energy range completely --- they produce little
contamination of the capture-induced NR ``signal.''

The \texttt{Geant4} particle transport code gets different results for the resulting NR spectra
from the capture process with silicon, germanium, and neon~\cite{Villano2022}. The results for
silicon are the closest but still have significant differences that may be experimentally relevant
to dark matter and \cevns\ studies. The most recent version of \texttt{Geant4} that was compared
to our spectral results is \texttt{v10.7.3} and comparisons are stored with our open-source
code~\cite{Villano2022}.

\section{\label{sec:calib_bknd}\texorpdfstring{Uses for Dark Matter and CE\MakeLowercase{\textnu{}}NS}{Uses for Dark Matter and CEnNS}}

Our major goals in understanding the neutron capture induced nuclear recoil spectra are (a) to use
these nuclear recoil events to enhance our understanding of low-energy nuclear recoils in
solids --- to provide excellent low-energy calibrations and (b) to compute the experimental
backgrounds for dark matter and \cevns\ searches.

In silicon and other materials there has not been consensus on how much ionization nuclear recoils
produce at low recoil energies. Typically, our theoretical guidance in the field comes from the
early Lindhard paper~\cite{osti_4701226} and associated work. On the other hand, measurements down
to the 100\,eV range seem like they may deviate from those predictions \emph{and} be marginally
consistent or  inconsistent with each
other~\cite{Izraelevitch_2017,PhysRevD.94.082007,PhysRevD.105.083014}. 

\begin{figure}[!htb]
   \includegraphics[width=\columnwidth]{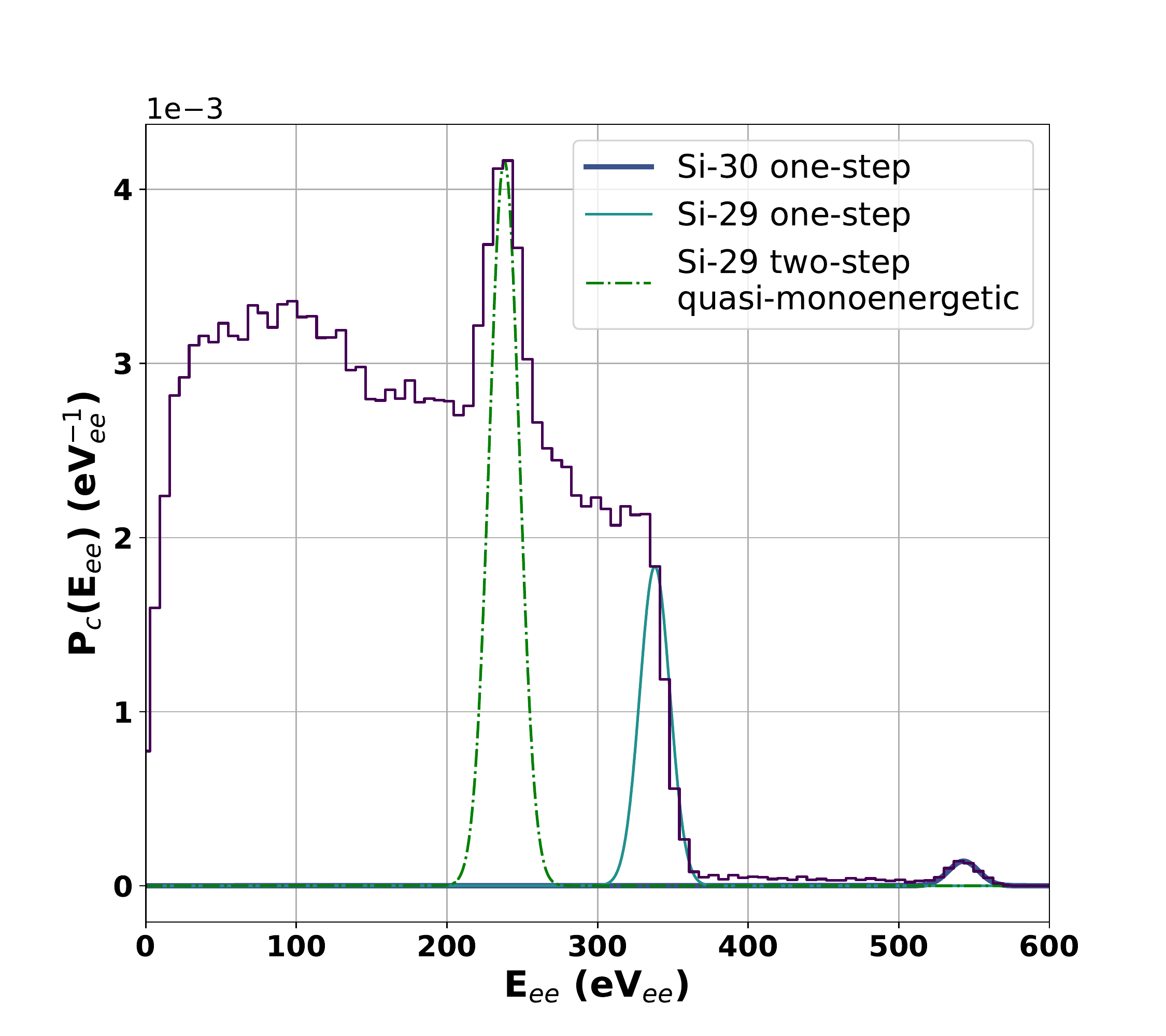}
   \caption{
       \label{fig:fullcapspec-Eee}(Color online) The NR capture spectrum
    corresponding to events that are resulting from neutron-capture. The three Gaussian peaks shown
    outline the particular cascades that have sharp signatures. The horizontal scale is in
    electron-equivalent energy (eV$_{\mathrm{ee}}$) which quantifies the amount of charge produced
    rather than the true recoil energy of the NR (which is higher). 95.63\% of all
    cascades are modeled using the data from Table~\ref{tab:acc_prob}. A nominal
    10\,eV$_{\mathrm{ee}}$ resolution has been applied.
}
\end{figure}

Exploring the detector response to NRs at low energies is therefore prudent and thermal neutron
induced captures provide an excellent venue for this --- as pointed out by the CRAB
Collaboration~\cite{Thulliez_2021} and others working with xenon for dark
matter~\cite{PhysRevD.106.032007}. The key features for thermal neutron induced captures are shown
in Fig.~\ref{fig:fullcapspec-Eee}. In that figure a nominal Lindhard ionization
yield~\cite{osti_4701226} is applied to the result shown in Fig.~\ref{fig:fullcapspec-resolution}. 

Calibrations using thermal neutron induced captures are superior to other styles of calibrations
that have been used: direct elastic neutron scattering~\cite{PhysRevD.42.3211}, photoneutron
sources~\cite{PhysRevD.105.122002,PhysRevD.94.122003}, and $^{252}$Cf sources~\cite{AGNESE201871}.
Figure~\ref{fig:fullcapspec-Eee} shows that the spectrum has sharp mono-energetic features that
are lacking in wide-band photoneutron or $^{252}$Cf sources. The spectrum extends down well below
100\,eV which is probably near the limit of elastic scattering sources. Using this technique is
also feasible \emph{in situ} because any neutron source will elevate the thermal neutron flux
during its deployment. Finally, if there is a case where exiting gammas can be measured in
coincidence, the direct-to-ground transitions provide \emph{directionally tagged} nuclear recoils.
This would lead to a heretofore unavailable NR directionality calibration, as pointed out
in~\cite{Thulliez_2021}.

A large enough thermal neutron flux could lead to meaningful backgrounds for low-mass dark matter
searches or \cevns\ measurements. The thermal neutron flux is typically not measured directly in
many experiments because of the difficulty in doing so. One measurement that does exist for
\cevns\ is from the MINER collaboration~\cite{AGNOLET201753} and is several orders of magnitude
higher than the accepted sea level environmental value, 4~cm$^{-2}$\,hr$^{-1}$~\cite{1263842}. We
have previously shown the effect of thermal neutrons on \cevns\ measurements in
detail~\cite{biffl2022critical}.

The SuperCDMS Soudan thermal neutron flux can be estimated from the germanium activation
lines~\cite{PhysRevD.99.062001} and is $\lesssim$\,7.2$\times$10$^{-2}$~cm$^{-2}$\,hr$^{-1}$\footnote{This
value was estimated by examining the published 10.37\,keV line rate over a period of a month
starting at least three half-lives after a three-day (maximum) Cf activation, see Fig.~1 of the
reference. At that point the rate appears to be close to consistent with flat. Some time
later the estimated flux would be about 0.3 times this value.}.

\begin{figure}[!htb]
   \includegraphics[width=\columnwidth]{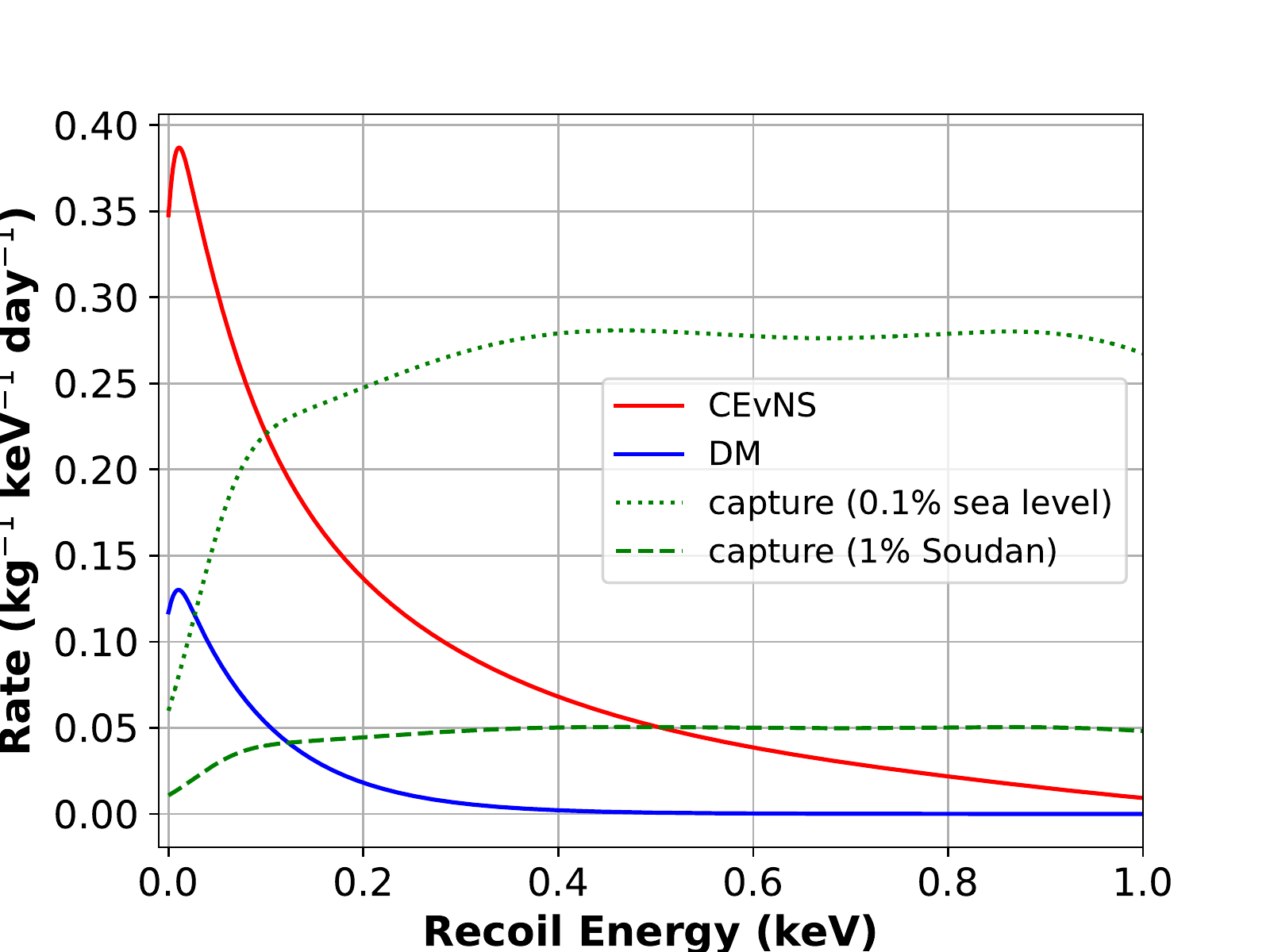}
   \caption{\label{fig:bkndcompare}(Color online) Comparison of the thermal neutron capture
induced NR spectrum in silicon with expected dark matter and \cevns\ signals in line
with currently available data. 
   }
\end{figure}

Figure~\ref{fig:bkndcompare} shows the comparison of the thermal neutron induced NR spectra at the
estimated flux levels without shielding with the dark matter and \cevns\ spectra. The capture
cross section is assumed to be 0.171~barns, from 2003 measurements at Brookhaven National
Lab~\cite{mughabghab2003capturesxn} (this value is also used in the EGAF (Evaluated Gamma
Activation File) database~\cite{egaf_paper}). This value is about 4\% higher than the value given
by evaluations a few years earlier at Los Alamos National Lab that are used in ENDF and the
JENDL~5 database~\cite{jendl5_paper} and the cross section measurements made by
Raman~\cite{PhysRevC.46.972}.

For a \cevns\ experiment with a 1\,MW reactor at a distance of 8\,m we arbitrarily compared a
thermal neutron flux of 0.1\% of the sea level value. We have used the Mueller spectrum for the
reactor anti-neutrinos~\cite{PhysRevC.83.054615}. The detector resolution function is assumed to
have a 10\,eV baseline that rises to 30\,eV at a recoil energy of 25\,eV. We use this form for the
energy-varying resolution: $\sqrt{\sigma_0^2 + AE_r}$; $E_r$ is the recoil energy and $\sigma_0$ and
$A$ are constants.

For the dark matter comparisons in Fig.~\ref{fig:bkndcompare} we used a 1\,GeV mass dark matter
particle with a cross section just below the limit produced in recent SuperCDMS
work~\cite{PhysRevD.99.062001,PhysRevD.97.022002}.

Figure~\ref{fig:bkndcompare} shows that both for dark matter searches and \cevns\ the spectral
overlap of thermal neutron capture induced NRs can interfere with measurements especially in cases
where the detector baseline resolutions are larger than 10\,eV --- true for all but the best modern
detectors. 

\section{\label{sec:conclusion}Conclusion}

We have carefully derived and simulated the spectrum of NRs following thermal neutron captures in
silicon. The spectra do not match the contemporary \texttt{Geant4} particle transport
code, indicating the details of decay-in-flight and atomic slowing-down are poorly modeled.

The level of thermal neutron fluxes that may be present in underground laboratories (mostly from
radiogenic sources in deep labs) is comparable to the contemporary rate limits on dark matter
scattering. Furthermore, in the \cevns\ venue the thermal neutron capture background could also
play an important role due to the proximity of some experiments to neutron-generating
reactors~\cite{PhysRevLett.129.211802,ANG2021165342}. In both of these situations the authors
recommend studies of the thermal neutron flux levels and taking the thermal neutron capture
background into account during data analysis.

\begin{acknowledgements}
We gratefully acknowledge support from the U.S. Department of Energy (DOE)
Office of High Energy Physics and from the National Science Foundation (NSF).
This work was supported by DOE Grant No. DE-SC0021364 and in part by NSF Grant
No.~2111090.  
\end{acknowledgements}

\clearpage
\bibliography{prdrefs_short.bib}
\bibliographystyle{apsrev4-1}
\clearpage 
\appendix
\section{\label{app:rearrange}Another perspective on equations}

In Sec. \ref{sec:twostep}, we give several equations that are constructed from the perspective of simulation. Below are the derivations, from the equations in Sec. \ref{sec:twostep}, of equivalent forms that give a more conceptual perspective. 

We use the following substitutions. Let $v_1 = (2(T_1-D_1)/M_A^*)^{1/2}$ be the velocity just before the second gamma's emission, $v_{CM} = \Delta/M_A$ be the velocity just after emission in the center-of-mass frame, and $E_1 = \frac{1}{2}M_A^*v_1^2$ and $E_CM = \frac{1}{2}M_Av_{CM}^2$ be the associated kinetic energies. 

 First, we will reorganize $g^0(D_1)$ from Eq.~(\ref{eq:gd1}):

\begin{equation}\label{eq:g0}\begin{aligned}
    g^0(D_1) & = \frac{v_1\ln 2}{v_1|a|t_{1/2}M_A^*\sqrt{2(T_1-D_1)/M_A^*}}\exp{\left(-\ln 2\frac{v_0-v_1}{|a|t_{1/2}}\right)} \\
    & = \frac{v_1\ln 2}{|a|t_{1/2}M_A^*v_1^2}\exp{\left(-\ln 2\frac{v_0-v_1}{|a|t_{1/2}}\right)} \\
    & = \frac{\ln 2}{2E_1}\frac{v_1}{|a|t_{1/2}}\exp\left(-\ln 2\frac{v_0-v_1}{|a|t_{1/2}}\right).\\
\end{aligned}\end{equation}

Next, we reorganize $T_2$, given by Eq.~(\ref{eq:d2minmax}):

\begin{widetext}\begin{equation}\label{eq:T2}\begin{aligned}
    T_2 & = \frac{1}{2}M_A\frac{\Delta^2}{M_A^2}\left[\frac{M_A^2M_A^*\sqrt{2(T_1-D_1)/M_A^*}^2}{M_A\Delta^2} +2\sqrt{M_AM_A^*}\sqrt{\frac{2(T_1-D_1)/M_A}{\Delta^2}}\cos \beta_{CM} + 1\right] \\
    & = \frac{1}{2}M_Av_{CM}^2\left[\frac{M_A^*v_1^2}{M_Av_{CM}^2} + 2\sqrt{\frac{M_A^*v_1^2}{M_Av_{CM}^2}}\cos \beta_{CM}+1\right] \\
    & = E_CM\left[\frac{E_1}{E_{CM}} + 2\sqrt{\frac{E_1}{E_{CM}}}\cos \beta_{CM} + 1\right]
\end{aligned}\end{equation}\end{widetext}

This new form for $T_2$ makes it clear that $T_2$ can reach zero when $E_1 = E_{CM}$ and $\beta = \pi$.

Finally, we reorganize $h(D_2|D_1)$ for the nonzero case, given by Eq.~(\ref{eq:h}):

\begin{equation}\label{eq:hnew}\begin{aligned}
    h(D_2|D_1) & = \frac{M_A}{2M_A\Delta}\sqrt{\frac{1/M_AM_A^*}{2(T_1-D_1)/M_A^*}} \\
    &= \frac{2}{4M_Av_{CM}}\sqrt{\frac{M_A}{M_A^*}}\frac{1}{v_1} \\
    &= \frac{2}{4M_Av_{CM}^2}\sqrt{\frac{M_Av_{CM}^2}{M_A^*v_1^2}} \\
    &= \frac{1}{4E_{CM}}\sqrt{\frac{E_{CM}}{E_1}}
\end{aligned}\end{equation}

\end{document}